\documentclass[aps,prl,showpacs,notitlepage,singlecolumn,superscriptaddress,11pt,floatfix,]{revtex4-1}
\usepackage{amsmath}
\usepackage{amssymb}
\usepackage{amsfonts}
\usepackage{stmaryrd}
\usepackage{wasysym}
\usepackage{graphicx}
\usepackage{multirow}
\usepackage{tabularx}
\usepackage{textcomp}
\usepackage{subfigure}
\usepackage{comment} 
\usepackage{url}
\usepackage[colorlinks=true,citecolor=blue,urlcolor=blue]{hyperref}
\usepackage{romannum}
\usepackage{mathtools}
\usepackage[utf8]{inputenc}
\usepackage{braket}
\usepackage{amsmath}
\usepackage{xcolor}
\usepackage{colortbl}
\usepackage{color,soul} 
\usepackage{bm}
\usepackage[normalem]{ulem}
\usepackage[utf8]{inputenc}
\usepackage{array}
\usepackage{makecell}
\usepackage{multirow}
\usepackage{enumitem}  
\hypersetup{colorlinks=true, urlcolor=blue, citecolor=cyan, pdfborder={0 0 0}}

\usepackage{fancyhdr}
\usepackage{pdfpages} 
\usepackage{pgffor} 
\makeatletter
\AtBeginDocument{\let\LS@rot\@undefined}
\makeatother
\def\supplementfilename{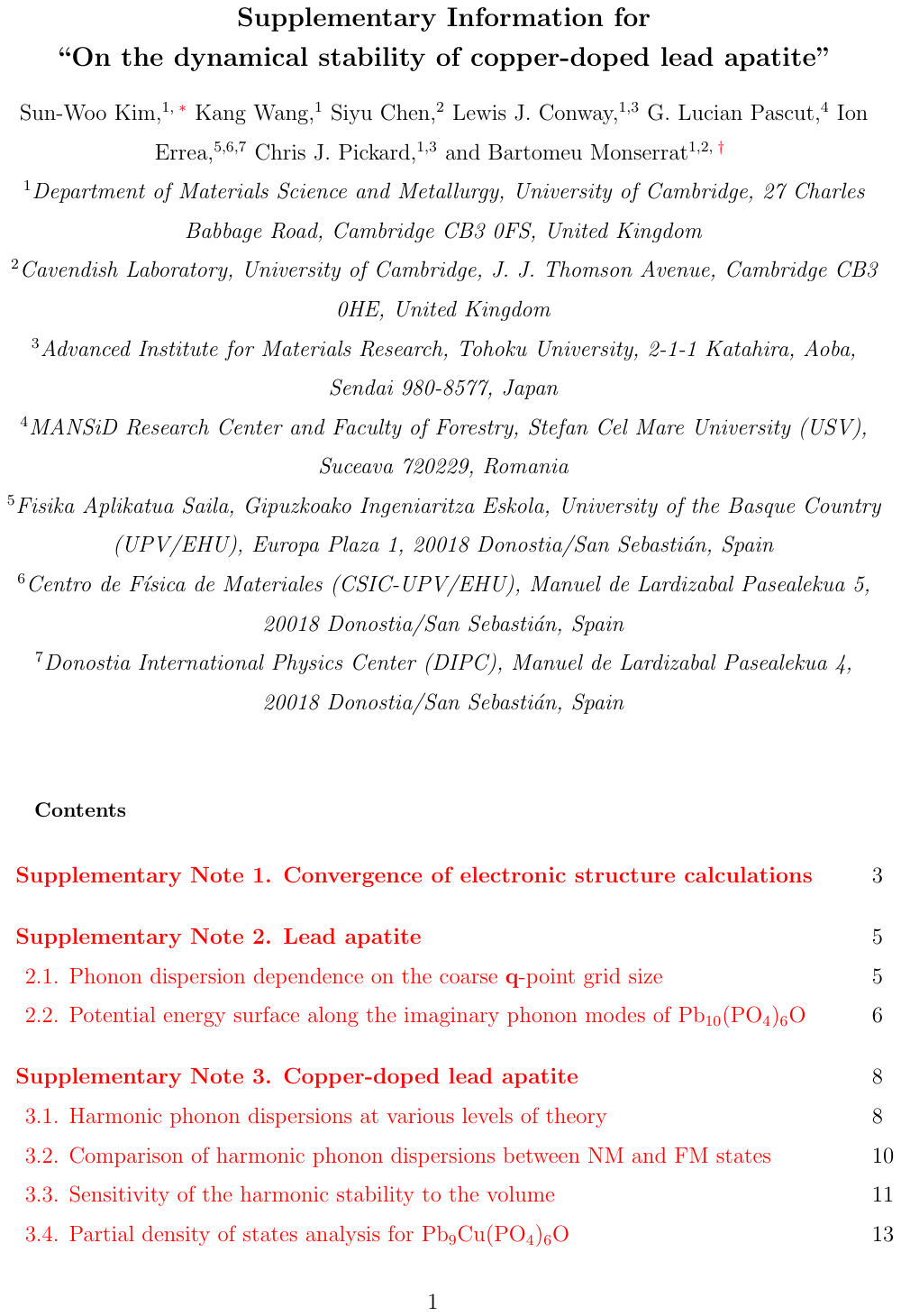}

\pdfximage{\supplementfilename}
\def\numbersupplementpages{\the\pdflastximagepages}

\newif\ifarXiv
\arXivtrue 

\begin{document}
\pagenumbering{arabic}

\title{On the dynamical stability of copper-doped lead apatite}
\maketitle
\begin{center}

{Sun-Woo Kim,$^{1,\;\textcolor{red}{*}}$ Kang Wang,$^{1}$ Siyu Chen,$^{2}$ Lewis J. Conway,$^{1,3}$ G. Lucian \\ Pascut,$^{4}$ Ion Errea,$^{5,6,7}$ Chris J.  Pickard,$^{1,3}$ and Bartomeu Monserrat$^{1,2,\;\textcolor{red}{\dagger}}$
}

\emph{$^{1}$Department of Materials Science and Metallurgy, University of Cambridge,\\ 27 Charles Babbage Road, Cambridge CB3 0FS, United Kingdom}\\
\emph{$^{2}$Cavendish Laboratory, University of Cambridge,\\ J. J. Thomson Avenue, Cambridge CB3 0HE, United Kingdom}\\
\emph{$^{3}$Advanced Institute for Materials Research, Tohoku University,\\ 2-1-1 Katahira, Aoba, Sendai 980-8577, Japan}\\
\emph{$^{4}$MANSiD Research Center and Faculty of Forestry,\\ Stefan Cel Mare University (USV),
Suceava 720229, Romania}\\
\emph{$^{5}$Fisika Aplikatua Saila, Gipuzkoako Ingeniaritza Eskola,\\ University of the Basque Country (UPV/EHU),\\ Europa Plaza 1, 20018 Donostia/San Sebastián, Spain}\\
\emph{$^{6}$Centro de Física de Materiales (CSIC-UPV/EHU),\\ Manuel de Lardizabal Pasealekua 5, 20018 Donostia/San Sebastián, Spain}\\
\emph{$^{7}$Donostia International Physics Center (DIPC),\\ Manuel de Lardizabal Pasealekua 4, 20018 Donostia/San Sebastián, Spain}\\

$^{\textcolor{red}{*}}$ \textup{email: \href{mailto:swk38@cam.ac.uk}{swk38@cam.ac.uk}}\\
$^{\textcolor{red}{\dagger}}$ \textup{email: \href{mailto:bm418@cam.ac.uk}{bm418@cam.ac.uk}}
\end{center}

\section{Abstract}
The recent claim of room temperature superconductivity in a copper-doped lead apatite compound, called LK-99, has sparked remarkable interest and controversy. Subsequent experiments have largely failed to reproduce the claimed superconductivity, while theoretical works have identified multiple key features including strong electronic correlation, structural instabilities, and dopability constraints. A puzzling claim of several recent theoretical studies is that both parent and copper-doped lead apatite structures are dynamically unstable at the harmonic level, questioning decades of experimental reports of the parent compound structures and the recently proposed copper-doped structures. In this work, we demonstrate that both parent and copper-doped lead apatite structures are dynamically stable at room temperature. Anharmonic phonon-phonon interactions play a key role in stabilizing some copper-doped phases, while most phases are largely stable even at the harmonic level. We also show that dynamical stability depends on both volume and correlation strength, suggesting controllable ways of exploring the copper-doped lead apatite structural phase diagram. Our results fully reconcile the theoretical description of the structures of both parent and copper-doped lead apatite with experiment.

\section{Introduction}

Copper-doped lead apatite {Pb}$_{10-x}${Cu}$_x$({PO}$_4$)$_6${O} with $0.9 < x < 1.1$, known as LK-99, has been recently claimed to exhibit superconductivity above room temperature and at ambient pressure\,\cite{LK99-1,LK99-2}. This remarkable claim is backed by magnetic (half-)levitation on a permanent magnet and by a sudden drop in resistivity at the claimed superconducting transition temperature. 
However, subsequent extensive experimental efforts by other groups have failed to confirm the superconductivity\,\cite{Kumar_2023_synthesis,Liu_2023_semiconducting,wu2023successful,Guo2023,Wang2023,zhang2023ferromagnetism,zhu2308first,timokhin2023synthesis,kumar2023absence,Liu_2023_phases,puphal2023single}. 
The magnetic half-levitation is reproduced in some insulating samples, where it is attributed to soft ferromagnetism\,\cite{Guo2023,Wang2023,zhang2023ferromagnetism}.
A plausible explanation for the sudden resistivity drop is provided by a first order phase transition of Cu$_2$S impurities\,\cite{zhu2308first,Jain2023}, which is further supported by the highly insulating nature of a single crystalline sample without Cu$_2$S impurities\,\cite{puphal2023single}. 

On the theoretical front, initial density functional theory calculations reported an electronic structure exhibiting relatively flat bands near the Fermi level for a simple model of copper-doped lead apatite\,\cite{griffin2023origin,si2023electronic,LAI202466,kurleto2023pb}. However, subsequent calculations showed that the inclusion of spin-orbit coupling or non-local correlations lead to an insulating electronic structure\,\cite{bai2023ferromagnetic,swift2023comment,pashov2023multiple}, a conclusion that is also reached with the inclusion of local correlations using dynamical mean-field theory\,\cite{korotin2023electronic,si2308pb10,yue2023correlated}. Different estimates of critical superconducting temperatures have so far delivered values significantly lower than room temperature\,\cite{oh2023s,witt2023no,paudyal2023implications}.

These state-of-the-art electronic structure calculations all assume a specific structural model as a starting point, often suggested by experiment. However, other theoretical works have questioned the suitability of these structural models both in terms of the thermodynamic feasibility of copper doping\,\cite{shen2023phase} or the dynamical stability of the experimentally proposed structures\,\cite{Bernevig_phonon,shen2023phase,hao2023first,liu2023symmetry,liu2023different,cabezas2023theoretical}. Indeed, one of the most basic quantities used to characterize a material is its dynamical stability. A dynamically stable structure corresponds to a local minimum of the potential (free) energy surface, and its phonon frequencies are real. A dynamically unstable structure corresponds to a saddle point of the potential (free) energy surface, and some of its phonon frequencies are imaginary with associated eigenvectors that encode atomic displacement patterns that lower the energy of the system. Only dynamically stable structures can represent real materials. Puzzlingly, recent computational works have claimed that the experimentally reported structures of the parent lead apatite\,\cite{Bernevig_phonon,shen2023phase,hao2023first,liu2023symmetry} and of the copper-doped lead apatite compounds\,\cite{Bernevig_phonon,shen2023phase,hao2023first,liu2023different,cabezas2023theoretical} are dynamically unstable at the harmonic level, which would imply that they cannot be the true structures of the materials underpinning LK-99, and would question the validity of most electronic structure calculations to date.

In this work, we demonstrate that both parent lead apatite and copper-doped lead apatite compounds are \textit{dynamically stable} at room temperature. The parent compounds are largely stable at the harmonic level, with some exhibiting very slight instabilities which are suppressed by quartic anharmonic terms. For the copper-doped compounds, dynamical stability at the harmonic level depends on the doping site, but even those that are dynamically unstable at the harmonic level are overall stable at room temperature with the inclusion of anharmonic phonon-phonon interactions. 

\section{Results}
\subsection{Lead apatite}

\begin{figure}[h]
 \centering
 \includegraphics[width=.6\textwidth]{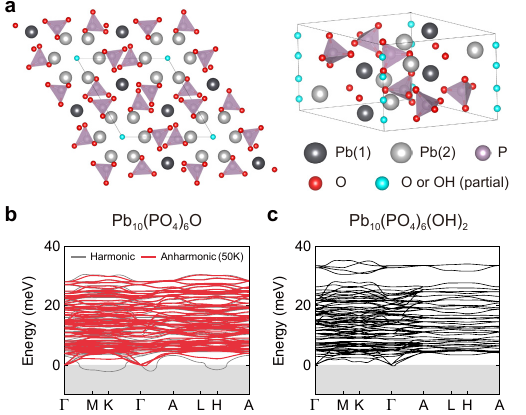} 
  \caption{ \textbf{Crystal structure and phonon dispersion of parent lead apatite.}
  \textbf{a.} Crystal structure of lead apatite Pb$_{10}$(PO$_4$)$_6$O or Pb$_{10}$(PO$_4$)$_6$(OH)$_2$. \textbf{b.} Harmonic and anharmonic (50\,K) phonon dispersions of Pb$_{10}$(PO$_4$)$_6$O. \textbf{c.} Harmonic phonon dispersion of Pb$_{10}$(PO$_4$)$_6$(OH)$_2$.
  }
  \label{fig:parent-phonon}%
\end{figure}

Lead apatite is a compound that was first experimentally reported over 70 years ago\,\cite{rooksby1952identification}. Figure\,\,\ref{fig:parent-phonon} depicts an example of lead apatite, with a hexagonal lattice (space group $P6_3/m$) and general formula Pb$_{10}$(PO$_4$)$_6$X$_2$, where X is either a halide atom or an OH group. The variant Pb$_{10}$(PO$_4$)$_6$O, which is claimed to be the parent structure of LK-99\,\cite{LK99-1,LK99-2}, has also been known experimentally for decades\,\cite{rooksby1952identification,merker1960oxypromorphit,merker1970lead,krivovichev2003crystal}. The X site corresponds to Wyckoff position $4e$, giving a multiplicity of four in the unit cell, but these sites are only partially filled with an occupation of $\frac{1}{2}$ for halide atoms and the OH group, and an occupation of $\frac{1}{4}$ for O. We consider two representative cases, Pb$_{10}$(PO$_4$)$_6$O and Pb$_{10}$(PO$_4$)$_6$(OH)$_2$, where the specific distribution of species on the X site results in space groups $P3$ and $P6_3$, respectively.

The phonon dispersion of the parent Pb$_{10}$(PO$_4$)$_6$O and Pb$_{10}$(PO$_4$)$_6$(OH)$_2$ compounds is shown in Fig.\,\ref{fig:parent-phonon}. At the harmonic level, Pb$_{10}$(PO$_4$)$_6$O exhibits imaginary phonon frequencies at the zone boundary points M and K of the $k_z=0$ plane, and at the zone boundary point H of the $k_z=\frac{\pi}{c}$ plane. However, the absolute values of these imaginary frequencies are less than $2.3$\,meV and the resulting anharmonic potentials have a dominant quartic term that strongly suppresses the instability to about $0.2$\,meV per formula unit (see Supplementary Figure 4). As a result, the calculation of self-consistent phonons including anharmonic interactions fully stabilizes the structure at the relatively low temperature of $50$\,K, and potentially lower. Earlier works reported that Pb$_{10}$(PO$_4$)$_6$O is dynamically unstable at the harmonic level\,\cite{Bernevig_phonon,shen2023phase,hao2023first}, a result we confirm, but our work further demonstrates that Pb$_{10}$(PO$_4$)$_6$O is overall dynamically stable at room temperature driven by higher-order anharmonic terms. 

Pb$_{10}$(PO$_4$)$_6$(OH)$_2$ is dynamically stable at the harmonic level. Earlier works reported that Pb$_{10}$(PO$_4$)$_6$(OH)$_2$ is dynamically unstable\,\cite{Bernevig_phonon,shen2023phase}, an opposite conclusion that we attribute to unconverged harmonic calculations (see Supplementary Figure 3). We note that to fully converge the harmonic calculations of Pb$_{10}$(PO$_4$)$_6$(OH)$_2$, the required coarse $\mathbf{q}$-point grid includes all of $\Gamma$, M, K, and A points, which can only be accomplished with a regular grid of minimum size $6\times6\times2$ or alternatively a nonuniform Farey grid\,\cite{farey_grid} of minimum size $(2\times2\times2)\cup(3\times3\times1)$. In our calculations, we use the latter as it is computationally more efficient. 
We highlight that our strategy, combining a nonuniform Farey grid\,\cite{farey_grid} with nondiagonal supercells\,\cite{nondiagonal_supercells,phonon_review}, offers a significant computational advantage in phonon calculations using the finite displacement method compared to the conventional diagonal supercell method with a regular grid used by most earlier works. Our approach drastically reduces the supercell sizes required to reach convergence, with the number of atoms decreasing from 3168 to only 132.

Overall, we find that both lead apatite compounds Pb$_{10}$(PO$_4$)$_6$O and Pb$_{10}$(PO$_4$)$_6$(OH)$_2$ are dynamically stable, a conclusion that is in full agreement with multiple experimental reports of the structure of lead apatite over the past 70 years\,\cite{rooksby1952identification,merker1960oxypromorphit,merker1970lead,krivovichev2003crystal,BRUCKNER1995209}.

\subsection{Copper doped lead apatite}

\begin{figure}[h]
 \centering
\includegraphics[width=\textwidth]{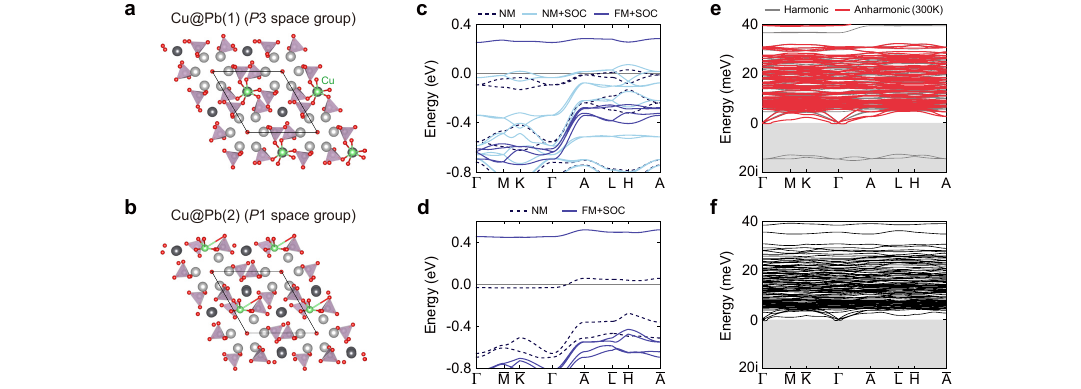} 
  \caption{
  \textbf{Electron band structure and phonon dispersion of Pb$_{9}$Cu(PO$_4$)$_6$O.}
  \textbf{a,b} Optimized crystal structures of Pb$_{9}$Cu(PO$_4$)$_6$O for copper doping at the \textbf{a} Pb(1) and \textbf{b} Pb(2) sites.
  \textbf{c,d} Electronic band structures of Pb$_{9}$Cu(PO$_4$)$_6$O for copper doping at the \textbf{c} Pb(1) and \textbf{d} Pb(2) sites. NM, FM, SOC refer to non-magnetic, ferromagnetic, and spin-orbit coupling, respectively. For doping at the Pb(2) site, the initial non-magnetic configuration converges to a ferromagnetic configuration in the presence of spin-orbit coupling. \textbf{e} Harmonic and anharmonic (300\,K) phonon dispersions of the Pb$_{9}$Cu(PO$_4$)$_6$O structure with $P3$ symmetry for doping at the Pb(1) site. \textbf{f} Harmonic phonon dispersion of the Pb$_{9}$Cu(PO$_4$)$_6$O structure with $P1$ symmetry for doping at the Pb(2) site.
  The phonon dispersions for the copper doped cases are obtained using the NM state without SOC (see SI for phonon dispersions of the FM state).
  The data are obtained with $U=3$\,eV on the copper $3d$ orbital.
  }
 \label{fig:doped-phonon}%
\end{figure}

The claim of room temperature superconductivity in LK-99 is based on copper doping of lead apatite, with copper replacing about 1 in 10 lead atoms leading to a Pb$_9$Cu(PO$_4$)$_6$O stoichiometry. There are two symmetrically distinct lead sites, labelled Pb(1) and Pb(2) in the literature (see Fig.\,\ref{fig:parent-phonon}\textbf{a}), and doping at these sites results in structures with the space groups $P3$ and $P1$, respectively (Fig.\,\ref{fig:doped-phonon}\textbf{a,b}). The original LK-99 work suggested that the doping site is Pb(1)\,\cite{LK99-1,LK99-2}, but subsequent experimental works have suggested that both Pb(1) and Pb(2) sites can be doped\,\cite{puphal2023single,Bernevig_phonon}. Computational works find that the relative energy between the two doping sites depends on the exchange-correlation functional and the magnitude of the Hubbard $U$ parameter used on the copper atom, with most choices favouring doping at the Pb(2) site, a prediction we confirm with our own calculations. For completeness, in this work we explore doping at both sites.

The electronic structures of Pb$_9$Cu(PO$_4$)$_6$O with copper on the Pb(1) and Pb(2) sites are shown in Fig.\,\ref{fig:doped-phonon}\textbf{c,d}. For doping at the Pb(1) site, a non-magnetic calculation leads to a metallic state in which the Fermi energy crosses four relatively flat bands (a pair of doubly-degenerate bands). Inclusion of spin-orbit coupling while maintaining the non-magnetic configuration leads to a splitting of the pair of doubly-degenerate bands, and the Fermi energy crosses a pair of singly-degenerate relatively flat bands. A calculation including spin-orbit coupling and allowing a non-zero magnetic moment leads to a ferromagnetic configuration in which the system is gapped. 
We note that the spin-orbit coupling is not essential for a gap opening in the presence of ferromagnetic ordering. The actual role of spin-orbit coupling, which lifts the orbital degeneracy, can be replicated by selecting a suitable level of theory, as evidenced by hybrid\,\cite{swift2023comment} or $GW$\,\cite{pashov2023multiple} calculations showcasing a gapped state in the presence of ferromagnetic ordering.
The ferromagnetic configuration is the most energetically favourable, but may not be directly relevant for room temperature experiments as single crystal measurements suggest the material is a non-magnetic insulator exhibiting a diamagnetic response with potentially a small ferromagnetic component\,\cite{puphal2023single}. Additionally, the ferromagnetic ordering may be an artifact of the DFT calculations, as dynamical mean-field theory calculations\,\cite{korotin2023electronic,si2308pb10,yue2023correlated} suggest a gap opens due to a Mott-like band splitting without the need of ferromagnetic ordering. For doping at the Pb(2) site, we also find a metallic state with a single band crossing the Fermi level in non-magnetic calculations, and a gapped state in ferromagnetic calculations. For both doping sites, we find that the phonon dispersion is only weakly affected by the level of electronic structure theory used (see Supplementary Notes 3.1 and 3.2), so the discussion below should be largely independent of the precise electronic structure of the system.
We hereafter present the phonon dispersions of the non-magnetic state calculated using PBEsol+$U$ without spin-orbit coupling.

The phonon dispersions of Pb$_9$Cu(PO$_4$)$_6$O with copper on the Pb(1) and Pb(2) sites are shown in Fig.\,\ref{fig:doped-phonon}\textbf{e,f}. Doping at the Pb(2) site leads to a dynamically stable structure at the harmonic level of theory. By contrast, doping at the Pb(1) site leads to a dynamically unstable structure at the harmonic level that exhibits two imaginary phonon branches of frequencies about $15i$\,meV across the entire Brillouin zone. This harmonic instability is present irrespective of the level of theory used, including a Hubbard $U$ parameter on the copper $d$ orbitals, spin-orbit coupling, and ferromagnetic ordering (see Supplementary Figure 5). Importantly, anharmonic phonon-phonon interactions strongly suppress the instability and the structure becomes dynamically stable at $300$\,K. We reach similar conclusions for copper doping of Pb$_{10}$(PO$_4$)$_6$(OH)$_2$ (see Supplementary Figure 6). Overall, copper-doped lead apatite is dynamically stable at room temperature for doping at either site.

\begin{figure}[h]
 \centering
 \includegraphics[width=\textwidth]{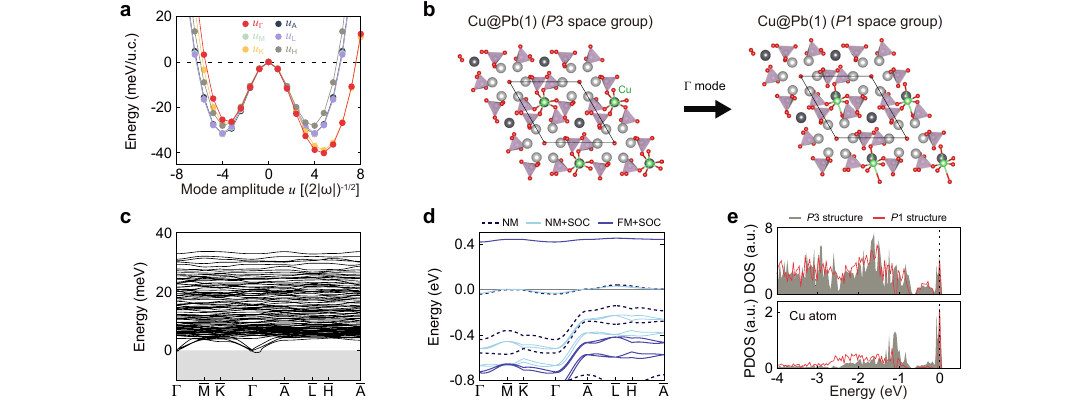} 
  \caption{
  \textbf{$\Gamma$-distorted $P1$ structure and its electron band structure and phonon dispersion.}
  \textbf{a.} 
  Potential energy surface along the imaginary phonon modes at high symmetry points in the Brillouin zone of the $P3$ structure for doping at the Pb(1) site (see its harmonic phonon dispersion in Fig. 2\textbf{e}).
 \textbf{b} Crystal structure distorted along the $\Gamma$ mode of the $P3$ structure.
  \textbf{c,d.} \textbf{c} Harmonic phonon dispersion and \textbf{d} electron band structure of the $\Gamma$-distorted $P1$ structure. 
 In \textbf{a-c}, the data are obtained using the NM state without SOC.
   \textbf{e.} Non-magnetic density of states (DOS) and Cu partial DOS of both the $P3$ and $\Gamma$-distorted $P1$ structures. See also partial DOS of other atoms in Supplementary Figure 10.
   The data are obtained with $U=3$\,eV on the copper $3d$ orbital.
  }
 \label{fig:pb1-doping}%
\end{figure}

The original paper claiming superconductivity in LK-99 suggested that copper doping of lead apatite occurs on the Pb(1) site\,\cite{LK99-1,LK99-2}. As the associated structure exhibits a dynamical instability at the harmonic level, we further explore its properties by considering the potential energy surface along the imaginary phonon modes at high symmetry points in the Brillouin zone (Fig.\,\ref{fig:pb1-doping}\textbf{a}). The dominant instability is driven by a $\Gamma$ point phonon mode, and fully relaxing the structure along this instability leads to a distinct structure of $P1$ symmetry (Fig.\,\ref{fig:pb1-doping}\textbf{b}; see also the link in Data availability statement for the optimized structure file).
We find that the $P1$ structure is dynamically stable at the harmonic level of theory (Fig.\,\ref{fig:pb1-doping}\textbf{c}), consistent with an earlier report\,\cite{cabezas2023theoretical}.
We ascribe the harmonic stability of the $P1$ structure to a downward shift of the occupied part of the density of states compared to the $P3$ structure (Fig.\,\ref{fig:pb1-doping}\textbf{e}), a trend similarly observed in the density of states of the harmonically stable Pb(2) doping case in Fig.\,\ref{fig:doped-phonon}\textbf{f} (see also Supplementary Figure 10 for the density of states of the Pb(2) doping case). This indicates that the harmonic stability is dominated by copper-derived orbitals.
The four bands (a pair of doubly-degenerate bands) that cross the Fermi level in the $P3$ structure (Fig.\,\ref{fig:doped-phonon}\textbf{c}) split under the distortion, such that the resultant $P1$ structure has a metallic state with a single doubly-degenerate band crossing the Fermi level in the non-magnetic configuration (Fig.\,\ref{fig:pb1-doping}\textbf{d}). The distorted $P1$ structure becomes an insulator in the presence of ferromagnetic ordering, which is consistent with previous results\,\cite{cabezas2023theoretical,georgescu2023cu}, similar to lead apatite with copper doping at the Pb(2) site (Fig.\,\ref{fig:doped-phonon}\textbf{d}). 

\begin{figure}[h]
 \centering
 \includegraphics[width=.65\textwidth]{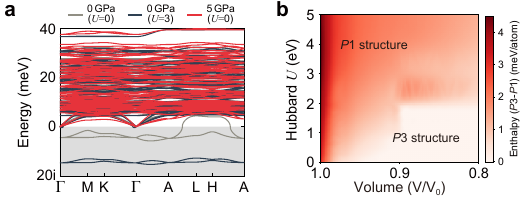} 
  \caption{
  \textbf{Volume-Hubbard $U$ phase diagram.}
  \textbf{a.} Representative harmonic phonon dispersions of the non-magnetic $P3$ structure doped at the Pb(1) site for different $U$ values and volumes. \textbf{b.} Volume-Hubbard $U$ phase diagram of Pb$_{9}$Cu(PO$_4$)$_6$O for doping at the Pb(1) site. The volume of the $P3$ structure is presented on the horizontal axis as a reference, corresponding to a pressure range of $0-25$ GPa. We find slightly larger volume changes in the $\Gamma$-distorted $P1$ structure. The color bar indicates the enthalpy difference between the $P3$ and $P1$ structures (in meV per atom). SOC is not considered in \textbf{a} and \textbf{b}.
  }
 \label{fig:phase-diagram}%
\end{figure}

Interestingly, the relative energy of the $P3$ structure compared to the $\Gamma$-distorted $P1$ structure is strongly dependent on both the volume and the electronic correlation strength as measured by the Hubbard $U$ parameter. Specifically, we find that harmonic instabilities favouring the $P1$ phase occur for large values of $U$ and large volumes, while the harmonic instability of the $P3$ phase completely disappears for small values of $U$ and small volumes. This is evident from the phonon dispersion of the $P3$ phase for different $U$ values and volumes (Fig.\,\ref{fig:phase-diagram}\textbf{a}) and from the enthalpy difference between $P3$ and $P1$ phases indicated by the colour bar in the phase diagram in Fig.\,\ref{fig:phase-diagram}\textbf{b}. These observations suggest that controlling volume, for example through hydrostatic pressure or strain, and controlling the degree of electronic correlation, for example by applying a gate voltage or doping, can be used to navigate the structural phase diagram of compounds based on lead apatite. Specifically, it may be possible to observe a temperature-driven structural phase transition between a low temperature $P1$ phase and high temperature $P3$ phase in a regime with a large harmonic dynamical instability. Finally, we note that electronic correlation beyond the static description provided by a Hubbard $U$ correction may play an important role on this phase diagram\,\cite{korotin2023electronic,si2308pb10,yue2023correlated}, so further work is required to fully characterise it. 

\section{Discussion}

Since the original claim of room temperature superconductivity in LK-99, seven phonon dispersion calculations have been reported in the literature. Of these, parent\,\cite{Bernevig_phonon,shen2023phase,hao2023first,liu2023symmetry} and copper-doped lead apatite with a $P3$ space group\,\cite{Bernevig_phonon,shen2023phase,hao2023first,liu2023different,cabezas2023theoretical} are claimed to be dynamically unstable at the harmonic level, while another work claims that the copper-doped lead apatite is dynamically stable at the harmonic level\,\cite{paudyal2023implications}.

We attribute these puzzling and contradictory conclusions about the dynamical stability of lead apatite to the complexity of harmonic phonon calculations in this system, with a unit cell containing at least $41$ atoms, and to the subtle interplay between volume, electronic correlation strength, and phonons. First, we find that fully converged phonon calculations for the parent compound Pb$_{10}$(PO$_4$)$_6$(OH)$_2$ require relatively large coarse $\mathbf{q}$-point grids, specifically incorporating the $\Gamma$, M, K, and A points. However, none of the previously reported calculations include all these points, and as a result they incorrectly conclude that this compound is dynamically unstable at the harmonic level. Second, we find that dynamical stability for the copper-doped compounds at the harmonic level depends on both the value of the Hubbard $U$ parameter and the volume of the system (Fig.\,\ref{fig:phase-diagram}; see also Supplementary Note 3.3). In this context, we rationalise the seemingly contradictory conclusions about dynamical stability of the copper-doped compounds by suggesting that different works use different volumes and different choices for the Hubbard $U$ parameter. 

Beyond clarifying the dynamical stability of lead apatite at the harmonic level, we have shown that anharmonic phonon-phonon interactions play a key role in stabilising multiple lead apatite compounds. Overall, our calculations indicate that both parent and copper-doped lead apatite compounds are dynamically stable at room temperature.

We believe that lead apatite is a nice example to illustrate the ability of state-of-the-art first principles methods to fully characterise a complex and experimentally relevant system. However, our work also demonstrates that reliable results and conclusions can only be reached with a careful consideration of convergence parameters, such as the size of the $\mathbf{q}$-point grid, and physical models, such as the inclusion of anharmonic phonon-phonon interactions.

We show that the experimentally suggested structures of lead apatite and copper-doped lead apatite are dynamically stable at room temperature. Most structures are dynamically stable at the harmonic level, but some key structures, including the structure claimed to be responsible for superconductivity at ambient conditions, only becomes dynamically stable with the inclusion of anharmonic phonon-phonon interactions. Our results resolve a puzzling suggestion by multiple earlier computational works that claimed that the experimentally reported structures of both parent and copper-doped lead apatite compounds were dynamically unstable, and fully reconcile the current experimental and theoretical description of the structure of lead apatite.

\section{Methods}

{\em Electronic structure calculations. -}
We perform density functional theory (DFT) calculations using the Vienna \textit{ab initio} simulation package {\sc vasp}\,\cite{VASP1,VASP2}, which implements the projector-augmented wave method~\cite{PAW}. We employ PAW pseudopotentials with valence configurations $5d^{10} 6s^2 6p^2$ for lead, $3d^{10} 4s^1$ for copper, $3s^{2} 3p^3$ for phosphorus, $2s^{2} 2p^4$ for oxygen, and $1s^{1}$ for hydrogen. For the exchange-correlation energy, we use both the generalized-gradient approximation functional of Perdew-Burke-Ernzerhof (PBE)\,\cite{PBE} and its modified version for solids (PBEsol)\,\cite{PBEsol}. We find that experimental lattice parameters agree well with those predicted by PBEsol, and the data presented in the main text has been obtained using PBEsol (see comparison between PBE and PBEsol in Supplementary Figure 5\textbf{a}).
An on-site Hubbard interaction $U$ is applied to the copper $3d$ orbitals based on the simplified rotationally invariant DFT+$U$ method by Dudarev and co-workers\,\cite{DFT+U}. We have checked that DFT+$U$ gives almost identical lattice parameters to DFT. Converged results are obtained with a kinetic energy cutoff for the plane wave basis of $600$\,eV and a $\mathbf{k}$-point grid of size $4\times4\times5$ and $6\times6\times8$ for the primitive cell of the parent and copper-doped lead apatite, respectively (see convergence test results in Supplementary Note 1). The geometry of the structures is optimised until all forces are below $0.01$\,eV per \AA\@ and the pressure is below $1$\,kbar. 

{\em Harmonic phonon calculations. -} We perform harmonic phonon calculations using the finite displacement method in conjunction with nondiagonal supercells\,\cite{nondiagonal_supercells,phonon_review}. We find that a coarse $\mathbf{q}$-point grid of size $2\times2\times2$ leads to converged phonon dispersions for the parent compound Pb$_{10}$(PO$_4$)$_6$O and the Cu-doped compounds. However, for the parent compound Pb$_{10}$(PO$_4$)$_6$(OH)$_2$, a converged calculation requires a minimum coarse $\mathbf{q}$-point grid including the high symmetry points $\Gamma$, M, K, and A, which we accomplish by means of a Farey nonuniform grid\,\cite{farey_grid} of size $(2\times2\times2)\cup(3\times3\times1)$. To evaluate the force derivatives, we use a three-point central formula with a finite displacement of $0.02$\,bohr. The underlying electronic structure calculations are performed using the same parameters as those described above. We have also cross-checked the phonon dispersions by performing additional calculations with 
{\sc castep}\,\cite{CASTEP} and {\sc Quantum Espresso}\,\cite{QE} within the finite displacement method and density functional perturbation theory, respectively (see Supplementary Figure 2).

{\em Anharmonic phonon calculations. -} We perform anharmonic phonon calculations using the stochastic self-consistent harmonic approximation (SSCHA)\,\cite{SSCHA1,SSCHA2,SSCHA3}, which accounts for anharmonic effects at both zero and finite temperature. The self-consistent harmonic approximation\,\cite{SSHA} is a quantum variational method on the free energy, and the variational minimization is performed with respect to a trial harmonic system. In its stochastic implementation, the forces on atoms are calculated in an ensemble of configurations drawn from the trial harmonic system. We use {\sc vasp} to perform electronic structure calculations using the same parameters as those described above, and consider configurations commensurate with a $2\times2\times2$ supercell. The number of configurations needed to converge the free energy Hessian is of the order of $4{,}000$ for the parent lead apatite structure and of the order of $8{,}000$ configurations for the copper-doped structure.

\section{Data availability}
The data that support the findings of this study are available within the paper and Supplementary Information. In particular, the optimized structure files can be found at \url{https://github.com/monserratlab/LK99_structures}.

\section{Code availability}
The {\sc VASP} code used in this study is a commercial electronic structure modeling software, available from \url{https://www.vasp.at}. The {\sc castep} code used in this study is open source for academic use, available from \url{http://www.castep.org/}. The {\sc Quantum Espresso} code used in this research is open source: \url{https://www.quantum-espresso.org/}.
The SSCHA code used in this research is open source: the SSCHA suite can be downloaded from \url{https://www.sscha.eu/} 

\section{Acknowledgments}

\begin{acknowledgments}
S.-W.K., K.W., and B.M. are supported by a UKRI Future Leaders Fellowship [MR/V023926/1]. B.M. also acknowledges support from the Gianna Angelopoulos Programme for Science, Technology, and Innovation, and from the Winton Programme for the Physics of Sustainability. S.C. acknowledges financial support from the Cambridge Trust and from the Winton Programme for the Physics of Sustainability. G.L.P. acknowledges funding from the Ministry of Research, Innovation, and Digitalisation within Program 1-Development of National Research and Development System, Subprogram 1.2-Institutional Performance-RDI Excellence Funding Projects, under contract no.\,10PFE/2021. I.E. acknowledges funding from the Department of Education, Universities and Research of the Eusko Jaurlaritza, and the University of the Basque Country UPV/EHU (Grant No. IT1527-22) and the Spanish  Ministerio de Ciencia e Innovación (Grant No. PID2022- 142861NA-I00). The computational resources were provided by the Cambridge Tier-2 system operated by the University of Cambridge Research Computing Service and funded by EPSRC [EP/P020259/1] and by the UK National Supercomputing Service ARCHER2, for which access was obtained via the UKCP consortium and funded by EPSRC [EP/X035891/1].

\textit{For the purpose of open access, the authors have applied a Creative Commons Attribution (CC BY) licence to any Author Accepted Manuscript version arising from this submission.}
\end{acknowledgments}

\section{Competing interests}
The authors declare no competing financial or non-financial interests.

\section{Author contributions}
S.-W.K., C.J.P., and B.M. conceived the study. S.-W.K. and B.M. planned and supervised the research. S.-W.K. and S.C. performed the DFT calculations, K.W. performed the SSCHA calculations. L.J.C., G.L.P., and I.E. provided advice on various aspects of the calculations. B.M. and S.-W.K. wrote the manuscript with input from all authors.

\def\bibsection{

\ifarXiv
    

\section*{Supplementary material}

## Supplementary Note 1. CONVERGENCE OF ELECTRONIC STRUCTURE CALCULATIONS

We have tested various convergence parameters for the electronic structure calculations underpinning the calculation of the phonon dispersions. We find that an energy cutoff of $600\,\text{eV}$ and a **k**-point grid size of $4 \times 4 \times 5$ are converged for the parent lead apatite $Pb_{10}(PO_4)_6O$, as illustrated in Fig. 1. For copper-doped lead apatite $Pb_9Cu(PO_4)_6O$ we find that a **k**-point grid size of $6 \times 6 \times 8$ leads to converged results.

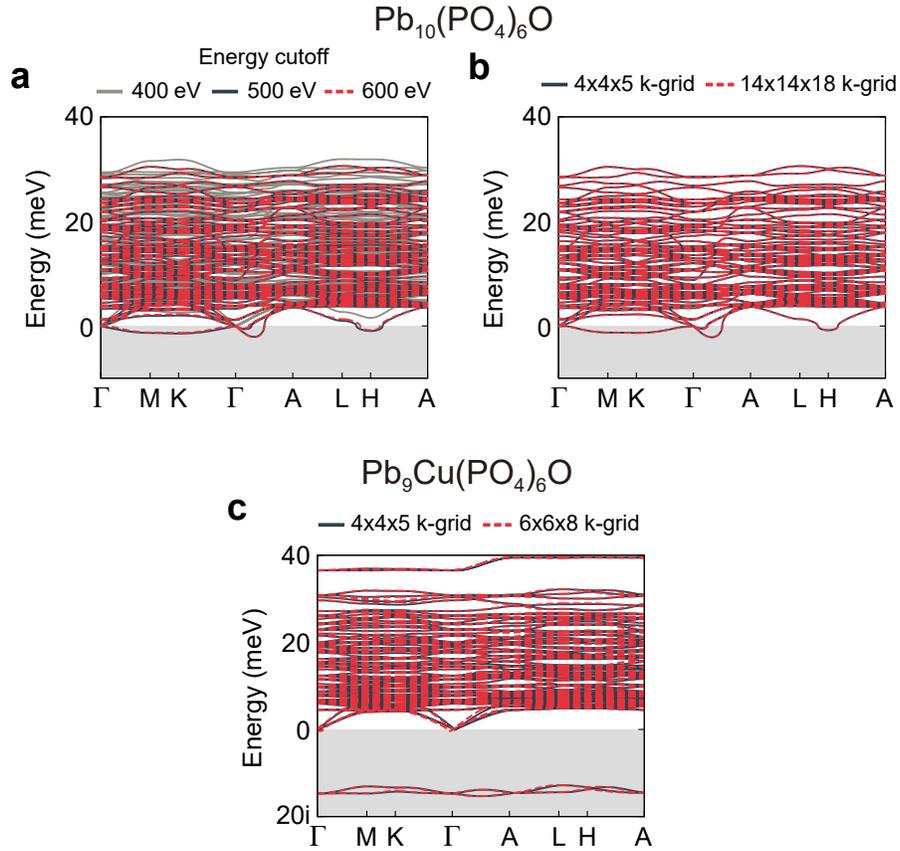

Supplementary Figure 1: **Convergence tests on the electronic structure calculations underpinning the phonon dispersions. a,b.** Convergence tests of phonon dispersions with respect to **a** energy cutoff and **b** k-point grid for $Pb_{10}(PO_4)_6O$. **c** The **k**-point grid convergence test on the phonon dispersion for $Pb_9Cu(PO_4)_6O$. The phonon dispersions for the copper doped case are obtained using a NM state without SOC. The calculations are performed using the PBE and PBEsol+$U$ ($U = 3\,\text{eV}$) methods for $Pb_{10}(PO_4)_6O$ and $Pb_9Cu(PO_4)_6O$, respectively.

We have also performed a cross-check of the phonon dispersion using three different codes: VASP [1, 2], CASTEP [3] and QUANTUM ESPRESSO [4]. We have confirmed that these codes

yield very similar results, as depicted in Fig. 2.

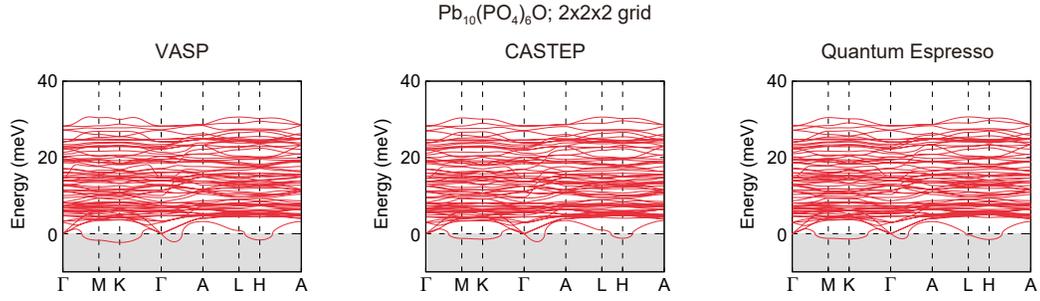

Supplementary Figure 2: **Cross-check of phonon dispersions. a-c.** Harmonic phonon dispersion of $Pb_{10}(PO_4)_6O$ calculated using **a** VASP [1, 2], **b** CASTEP [3], and **c** QUANTUM ESPRESSO [4]. The VASP and CASTEP calculations are done using the finite displacement method, while the QUANTUM ESPRESSO calculations are done using density functional perturbation theory. The PBEsol exchange-correlation functional and a $2 \times 2 \times 2$ **q**-point grid are used.

**Supplementary Note 2. LEAD APATITE**

### 2.1. Phonon dispersion dependence on the coarse q-point grid size

Figure 3 shows the phonon dispersions of $Pb_{10}(PO_4)_6O$ (top) and $Pb_{10}(PO_4)_6(OH)_2$ (bottom) for different choices of coarse **q**-point grid size. For $Pb_{10}(PO_4)_6O$, a qualitatively correct phonon dispersion is obtained with a coarse grid of size $2 \times 2 \times 2$, consistent with a previous study [5]. By employing larger grid sizes, we confirm that the imaginary frequencies at the K, M, and H points are physical rather than an artifact of Fourier interpolation.

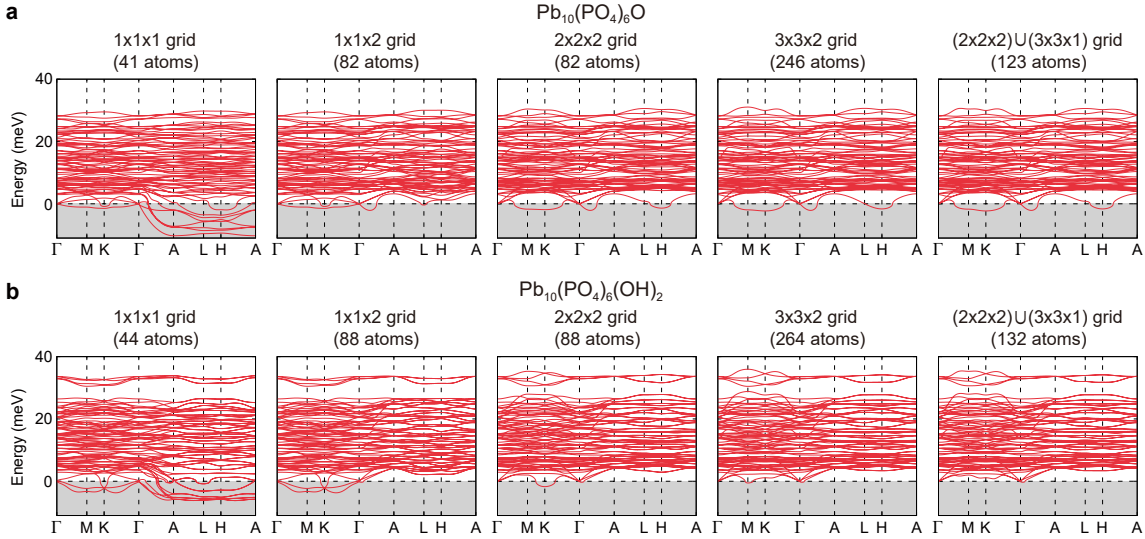

Supplementary Figure 3: **Coarse q-point grid dependence of the harmonic phonon dispersions. a,b.** Harmonic phonon dispersions of **a** $Pb_{10}(PO_4)_6O$ and **b** $Pb_{10}(PO_4)_6(OH)_2$ using different coarse **q**-point grids. The maximum number of atoms used in each **q**-point grid within the finite displacement method in conjunction with nondiagonal supercells [6, 7] is specified. The calculations are performed using the PBEsol exchange-correlation functional.

For $Pb_{10}(PO_4)_6(OH)_2$, a qualitatively correct phonon dispersion is only obtained when all of Γ, M, K, and A points are included in the coarse **q**-point grid, which can only be accomplished with a regular grid of minimum size $6 \times 6 \times 2$ or alternatively a non-uniform Farey grid [8] of minimum size $(2 \times 2 \times 2) \cup (3 \times 3 \times 1)$. We use the latter as it is computationally more efficient: within the nondiagonal supercell formalism [6, 7], it requires calculations using supercells with up to 132 atoms, compared to up to the 264 atoms required for the regular $6 \times 6 \times 2$ grid. We also highlight that our strategy, employing the nonuniform Farey grid with nondiagonal supercells, offers significant advantages in phonon calculations compared to

5

the conventional diagonal supercell method with a regular grid. This approach drastically reduces the maximum number of atoms required for a supercell, from 3168 to only 132, enabling us to obtain the converged phonon dispersion for $Pb_{10}(PO_4)_6(OH)_2$.

We note that all earlier phonon calculations for the $Pb_{10}(PO_4)_6(OH)_2$ compound in the literature use coarse **q**-point grids of sizes $1 \times 1 \times 1$ [9] or $1 \times 1 \times 2$ [10], and the imaginary phonon modes observed in these calculations are not physical but instead an artifact of Fourier interpolation caused by unconverged calculations. Indeed, the phonon dispersions reported in these works coincide with the corresponding unconverged calculations depicted in Fig. 3b.

### 2.2. Potential energy surface along the imaginary phonon modes of $Pb_{10}(PO_4)_6O$

The harmonic phonon dispersion of $Pb_{10}(PO_4)_6O$ exhibits imaginary frequencies at the M, K, and H points of the Brillouin zone, as depicted in the left panel of Fig. 4. The absolute values of these imaginary frequencies are below 2.3 meV, suggesting that the system is only marginally unstable at the harmonic level. To confirm this, we calculate the potential energy surface by displacing the atoms along the eigenvectors associated with the three imaginary modes, resulting in the double well potentials shown in the right panel of Fig. 4. The anharmonic potentials have a dominant quartic term that suppresses the instability to a maximum of 0.2 meV per formula unit for the K point, with smaller instabilities for the M and H points. By comparison, the thermal energy associated with room temperature is about 26 meV, suggesting that these shallow double well potentials can be overcome by thermally-induced anharmonic vibrations. We confirm this in the main text by performing self-consistent harmonic calculations at 50 K that deliver a dynamically stable phonon dispersion. The structure is likely dynamically stable at even lower temperatures, possibly at 0 K where it would be stabilized by quantum fluctuations.

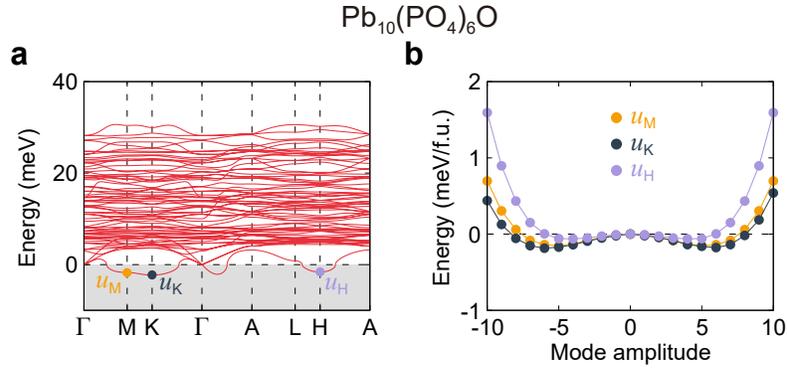

Supplementary Figure 4: **Potential energy surface of $Pb_{10}(PO_4)_6O$. a.** Harmonic phonon dispersion of $Pb_{10}(PO_4)_6O$ calculated using a Farey nonuniform grid of size $(2\times2\times2)\cup(3\times3\times1)$ which explicitly includes the K, M, and H points. **b.** Potential energy surface along the imaginary phonon modes at high symmetry points in the Brillouin zone. The calculations are performed using the PBEsol exchange-correlation functional.

**Supplementary Note 3. COPPER-DOPED LEAD APATITE**

### 3.1. Harmonic phonon dispersions at various levels of theory

The results regarding the dynamical stability of copper-doped lead apatite presented in the main text are robust against the level of theory used to describe the electronic structure of the system, as summarized in Fig. 5 for $Pb_9Cu(PO_4)_6O$. Panel a shows a comparison between the PBEsol (used in the main text) and PBE exchange-correlation functionals, which give qualitatively similar phonon dispersions and in particular they give similar imaginary branches. Panel b shows a comparison of the phonon dispersion for different values of Hubbard $U$ applied on the copper $3d$ orbital. In all cases there is an imaginary phonon branch, but the absolute value of the associated imaginary frequency increases with increasing $U$. We report results using $U = 3\,\text{eV}$ in the main text. For clarity, the evolution with $U$ of the $\Gamma$-point phonon imaginary frequency is detailed in panel c. Finally, panel d shows the imaginary frequency of the $\Gamma$-point phonon comparing the non-magnetic, non-magnetic with spin-orbit coupling, and ferromagnetic with spin-orbit coupling results. In all cases, the $\Gamma$-point phonon frequency is imaginary, although the magnitude changes by about $10\,\text{meV}$ depending on the level of theory. Overall, the harmonic dynamical instability of $Pb_9Cu(PO_4)_6O$ is robust against the level of electronic structure theory used.

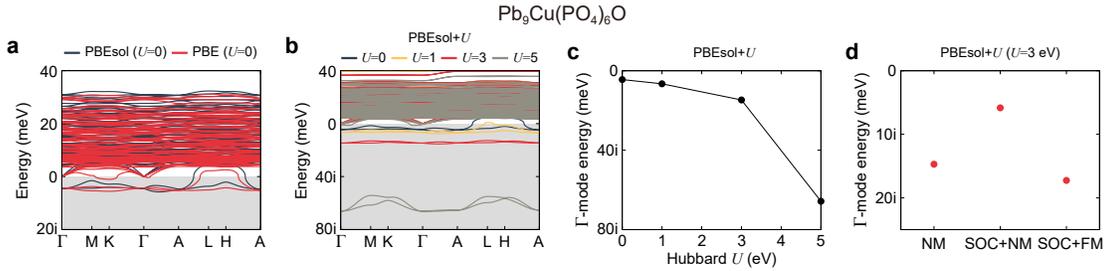

Supplementary Figure 5: **Harmonic phonon dispersion of $Pb_9Cu(PO_4)_6O$ at various levels of theory. a.** Harmonic phonon dispersion obtained using the PBE and PBEsol exchange-correlation functionals. **b.** Hubbard $U$ dependence of harmonic phonon dispersions, with $U$ applied to the copper $3d$ orbital. **c.** The lowest imaginary $\Gamma$-mode energy in the harmonic phonon dispersion as a function of Hubbard $U$. In **a-c**, the phonon dispersion and the $\Gamma$-mode energy are obtained using a NM state without SOC. **d.** Spin-orbit coupling and ferromagnetism effects on the lowest imaginary $\Gamma$-mode energy.

The phonon dispersion of the variant copper-doped apatite $Pb_{10}(PO_4)_6(OH)_2$ exhibits a similar behavior to that of $Pb_9Cu(PO_4)_6O$, as depicted in Fig. 6. In particular, we find a

8

similar dependence on the Hubbard $U$ parameter, in which the magnitude of the imaginary frequency increases with increasing $U$, and a similarly robust imaginary frequency with the inclusion of spin-orbit coupling and ferromagnetic order.

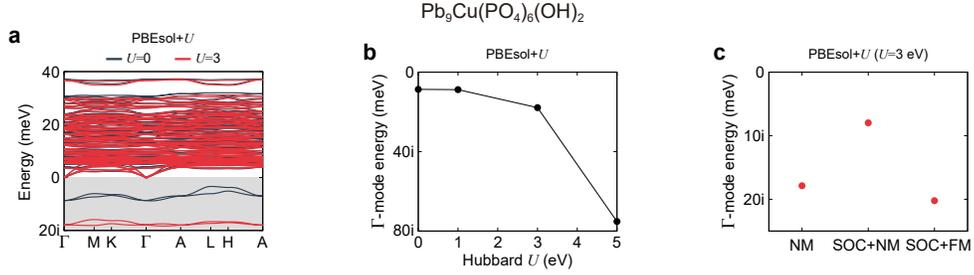

Supplementary Figure 6: **Harmonic phonon dispersion of $Pb_9Cu(PO_4)_6(OH)_2$ at various levels of theory. a.** Hubbard $U$ dependence of harmonic phonon band dispersion with $U$ applied to the copper $3d$ orbital. **b.** The lowest imaginary $\Gamma$-mode energy in the harmonic phonon dispersion as a function of Hubbard $U$. In **a-b**, the phonon dispersion and the $\Gamma$-mode energy are obtained using a NM state without SOC. **c.** Spin-orbit coupling and ferromagnetism effects on the lowest imaginary $\Gamma$-mode energy.

## 3.2. Comparison of harmonic phonon dispersions between NM and FM states

In the main text, we present the phonon dispersions of the non-magnetic state. This is because (i) the ferromagnetic state is not consistent with experiment, and (ii) the dynamical stability properties of lead apatite are largely independent of the precise electronic structure of the system, as discussed in Section 3.1 above. In addition, in this section, we provide both electronic structures and phonon dispersions of both non-magnetic and ferromagnetic states with and without spin-orbit coupling, aiming for a full picture of the system. We find harmonic instabilities in both non-magnetic and ferromagnetic states (see Fig. 7), showing marginal spin-phonon coupling.

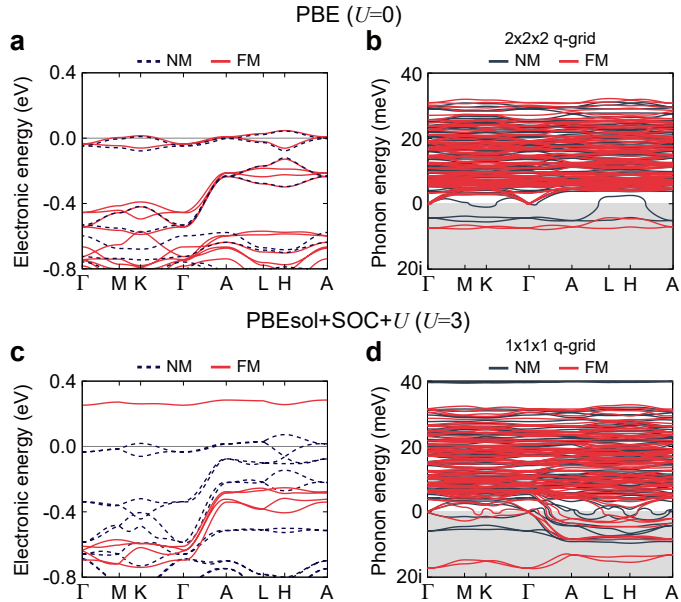

Supplementary Figure 7: **Comparison of electron band structure and phonon dispersion between non-magnetic and ferromagnetic states at various levels of theory for $Pb_9Cu(PO_4)_6O$.** **a** Electron band structure and **b** phonon dispersion of non-magnetic (NM) and ferromagnetic (FM) states obtained using the PBE functional. **c** Electron band structure and **d** phonon dispersion of NM and FM states obtained using the PBEsol+$U$ functional with spin-orbit coupling (SOC). In **d**, the phonon spectra is obtained using a $1 \times 1 \times 1$ **q**-point grid, which implies that only the phonons at the $\Gamma$ point are directly calculated. This is sufficient to investigate the $\Gamma$ point instability which is the primary instability in this system as discussed in Fig. 3 of the main text.

### 3.3. Sensitivity of the harmonic stability to the volume

In this section, we investigate the volume dependence of the harmonic stability of $Pb_9Cu(PO_4)_6O$ with copper on the Pb(1) site. This is motivated by the stable harmonic phonon dispersion reported in Ref. [11], which appears to be inconsistent with the results reported in other works [9, 10, 12–14].

We calculate PBE phonon dispersions using both our optimized volume of $619.29\,\text{Å}^3$ and the larger volume of $637.94\,\text{Å}^3$ used in Ref. [11]. We confirm the results from Ref. [11] and find that the larger volume gives rise to dynamical stability. However, we note that such dynamical stability only exists at a relatively narrow range of volume values as shown in Fig. 8c, indicating that the harmonic stability reported in Ref. [11] is a fortuitous result arising from the specific volume used, as most other volumes lead to harmonic instabilities. We therefore explain the apparent discrepancy between the phonon dispersion reported in Ref. [11] compared to those reported elsewhere [9, 10, 12–14], and attribute the differences to different volumes in the calculations.

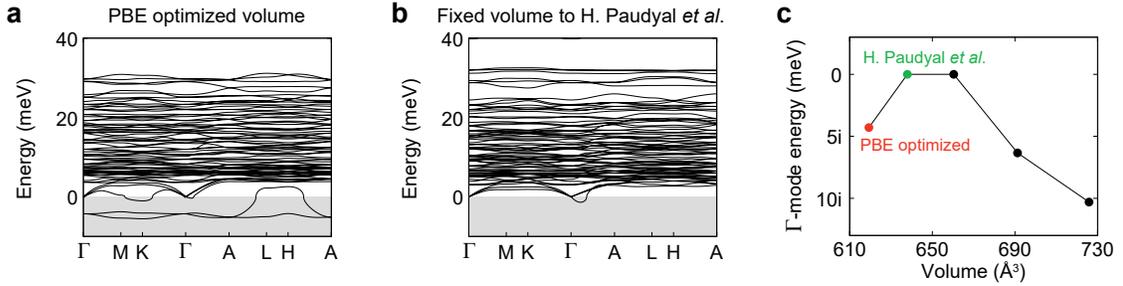

Supplementary Figure 8: **Volume dependence of phonon dispersion and harmonic instability of $Pb_9Cu(PO_4)_6O$. a,b** Harmonic phonon dispersions of the $P3$ structure for **a** the optimized volume of $619.29\,\text{Å}^3$ and **b** for the volume of $637.94\,\text{Å}^3$ used by Paudyal and co-workers [11]. **c** The lowest $\Gamma$-mode energy in the harmonic phonon dispersion as a function of volume. The calculations are performed using VASP with the PBE functional.

We also calculate the enthalpy difference between the $P3$ structure and $\Gamma$-distorted $P1$ structure under both negative and positive pressure (Fig. 9) to explore a wider range of volumes compared to those reported in the main text. At negative pressure, which results in a larger volume (where $-0.5\,\text{GPa}$ corresponds to the volume in Ref.[11]), the enthalpy difference $H(P_3) - H(P_1)$ increases. This clearly shows that $P1$ structure is more stable at larger volumes, as discussed in the main text.

Finally, remember that the $P1$ structure originally arises from an instability of the $P3$ structure at the $\Gamma$ point. The fact that this instability disappears for a narrow volume range (Fig. 8c) over which the $P1$ structure has a lower energy compared to the $P3$ structure suggests that an energy barrier develops between the two structures, and they are no longer connected by a dynamical instability. Interestingly, Fig. 8c also shows that a distinct instability appears for the $P3$ structure for even larger volumes.

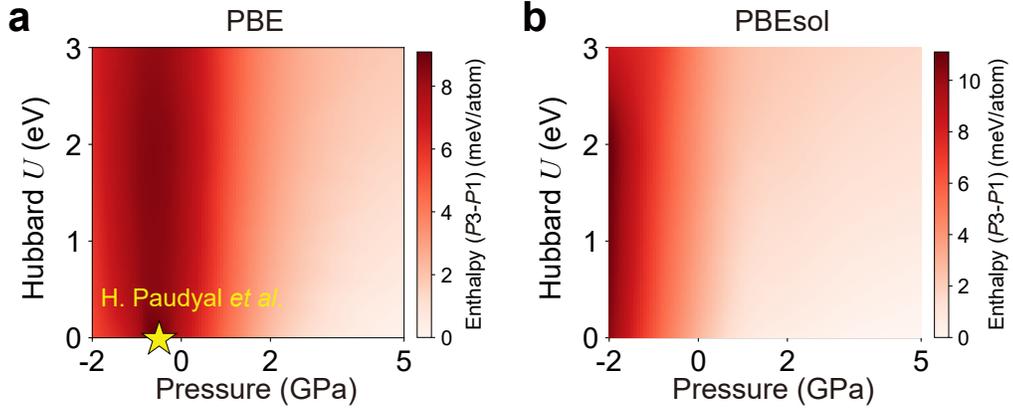

Supplementary Figure 9: **Volume-Hubbard $U$ phase diagram of $Pb_9Cu(PO_4)_6O$. a,b** Volume-Hubbard $U$ phase diagram obtained using **a** the PBE+$U$ functional and **b** the PBEsol+$U$ functional. In **a**, the yellow star indicates the pressure value ($-0.5$ GPa with $U = 0$) that gives the volume used in Ref. [11]. Both PBE+$U$ and PBEsol+$U$ results consistently show that the $P3$ structure is only stable over the positive pressure regime, corresponding to smaller volumes. The color bar quantifies the enthalpy difference between the $P3$ and $P1$ structures in meV per atom.

### 3.4. Partial density of states analysis for $Pb_9Cu(PO_4)_6O$

The $P3$ $Pb_9Cu(PO_4)_6O$ structure is dynamically unstable at the harmonic level, and its imaginary frequencies drive it to a lower-energy lower-symmetry $P1$ structure, which is dynamically stable. In the main text, we attribute the lower energy of the $P1$ structure compared to the $P3$ structure to a downward shift of the occupied part of the density of states (DOS) dominated by copper-derived orbitals, as illustrated in Fig. 10. We find a similar downward shift of the density of states when comparing the $P1$ structure doped at the Pb(2) site, as also illustrated in Fig. 10. Overall, this analysis suggests that the dynamical stability of the $P1$ phases with doping at either the Pb(1) or Pb(2) sites is driven by the relative energy of the copper states.

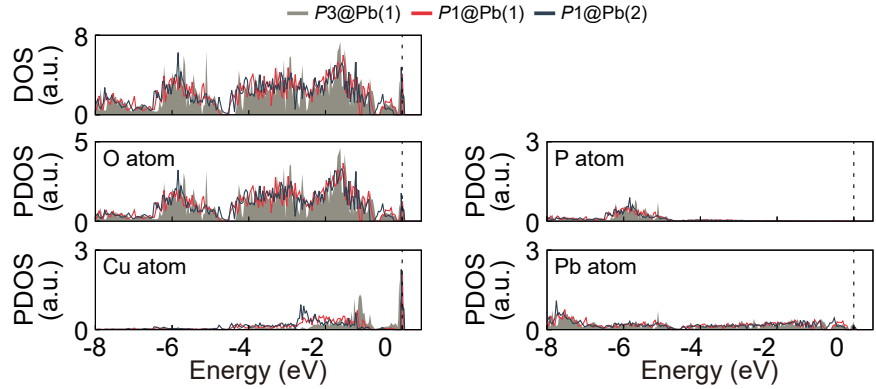

Supplementary Figure 10: **Origin of the lower energy of the $P1$ structures.** Non-magnetic DOS and atom-projected partial DOS of the $P3$ and $\Gamma$-distorted $P1$ structures for doping at the Pb(1) site [labeled as $P3$@Pb(1) and $P1$@Pb(1), respectively], and the $P1$ structure arising from doping at the Pb(2) site [$P1$@Pb(2)]. The data are obtained with PBEsol+$U$ with $U = 3$ eV.

13

### 3.5. Harmonic analysis for $Pb_9Cu(PO_4)_6(OH)_2$

In the main text, we have explored the potential energy surface along the imaginary phonon modes for $Pb_9Cu(PO_4)_6O$. Here, we perform a similar analysis for $Pb_9Cu(PO_4)_6(OH)_2$ and find similar results. The dominant instability is driven by a $\Gamma$ point phonon mode (Fig. 11a), and fully relaxing the structure along this instability leads to a new structure of $P1$ symmetry, which is dynamically stable at the harmonic level of theory (Fig. 11b).

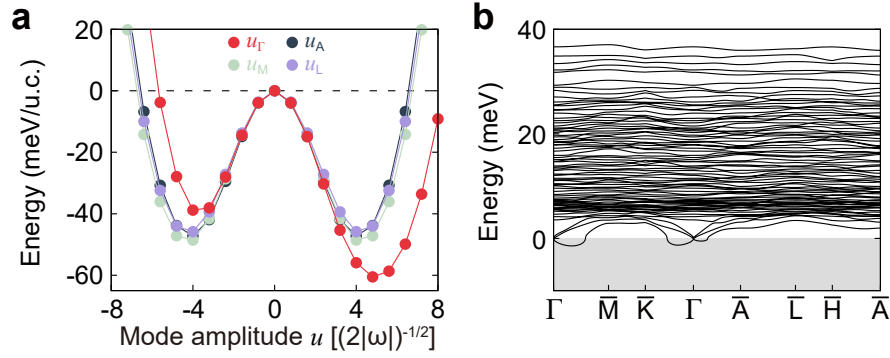

Supplementary Figure 11: **$\Gamma$-distorted $P1$ structure of $Pb_9Cu(PO_4)_6(OH)_2$ and its harmonic stability. a.** Potential energy surface along the imaginary phonon modes at high symmetry points in the Brillouin zone of the $P3$ structure for doping at the Pb(1). **b** Harmonic phonon dispersion of the $P1$ structure distorted along the $\Gamma$ mode of the $P3$ structure. The data are obtained with PBEsol+$U$ ($U = 3$ eV) for the NM state without SOC.

14


\fi

\end{document}